# Engineering quantum dots for electrical control of the fine structure splitting

M. A. Pooley[1,2], A. J. Bennett[1,◊], I. Farrer[2], D. A. Ritchie[2] and A. J. Shields[1]

1. Toshiba Research Europe Limited, Cambridge Research Laboratory, 208 Science Park, Milton Road, Cambridge, CB4 0GZ, U. K.

2. Cavendish Laboratory, Cambridge University, J. J. Thomson Avenue, Cambridge, CB3 0HE, U. K.

ABSTRACT We have studied the variation in fine-structure splitting (FSS) under application of vertical electric field in a range of quantum dots grown by different methods. In each sample we confirm that this energy splitting changes linearly over the field range we can access. We conclude that this linear tuning is a general feature of self-assembled quantum dots, observed under different growth conditions, emission wavelengths and in different material systems. Statistical measurements of characteristic parameters such as emission energy, Stark shift and FSS tuning are presented which may provide a guide for future attempts to increase the yield of quantum dots that can be tuned to a minimal value of FSS with vertical electric field.

---

◊ Corresponding author: e-mail: anthony.bennett@crl.toshiba.co.uk

The optically active exciton state, $|X\rangle$, has two spin-eigenstates $|X_H\rangle$ and $|X_V\rangle$ given by the two possible spin configurations of the constituent electron and hole. In general, the exciton state is time-dependent and can be expressed as a superposition of these two eigenstates such that

$$|X\rangle = \alpha_H |X_H\rangle + \exp\left(\frac{i|s|t}{\hbar}\right) \alpha_V |X_V\rangle, \tag{1}$$

where $\alpha_H^2 + \alpha_V^2 = 1$, and $s$ is the energy difference between the two linear eigenstates. This energy difference, $s$, is known as the fine-structure splitting (FSS), and is the key parameter that determines the character of entangled photon pairs emitted by the neutral biexciton cascade. The spin of the exciton state is mapped onto the polarization of photons emitted by this cascade, such that the resulting two-photon wavefunction is described by

$$|\Psi\rangle = \frac{1}{\sqrt{2}}\left(|HH\rangle + \exp\left(\frac{i|s|\tau_X}{\hbar}\right)|VV\rangle\right), \tag{2}$$

where $|HH\rangle$ ($|VV\rangle$) corresponds to both photons being emitted via state $|X_H\rangle$ ($|X_V\rangle$) with horizontal (vertical) polarization in the laboratory frame, and $\tau_X$ is the time that the system spends in state $|X\rangle$ between the two emission events. As seen in equation (2), a finite $|s|$ causes the polarization of the entangled photon pairs to oscillate, as a function of $\tau_X$, with a frequency of $|s|/\hbar$. In practice, if $|s| < 2\mu eV$ it is possible for standard avalanche photo-diodes, with a time resolution of 0.5 ns, to detect time-invariant entangled states with high fidelity to $|\Psi^+\rangle = \frac{1}{\sqrt{2}}(|HH\rangle + |VV\rangle)$.[1,2] However, for larger $|s|$ the phase accumulated by the intermediate exciton state varies too rapidly to be resolved and emission from the cascade appears to be only classically correlated[3,4]. For this reason there is much interest in reducing the FSS to as small a value as possible to allow the time-averaged emission to be entangled.

Numerous mechanisms have been demonstrated for tuning the FSS, but here we focus on the use of a vertical electric field, $F$. This is one of the most practical methods and has already led to the generation of entangled photon pairs[5] and the dynamic control of stored exciton spin-states[6]. In addition, it has recently been demonstrated that, in conjunction with an applied strain field, a vertical electric field can allow universal elimination of the FSS in all QDs within certain samples[7].

It has been shown that, for self-assembled QDs grown in a single deposition of InAs, the tuning rate of the FSS with electric field, $\gamma = ds/dF$, is constant[5]. To ensure photoluminescence is visible across a large range of electric fields we embed the QDs inside a diode with barriers on either side of the active region. This structure allows large fields to be applied without quenching the optical emission[8]. This device design has been used previously to vary the FSS of both InAs/GaAs QDs grown in a single deposition[5] and GaAs/AlAs QDs[9]. In this work, we extend this observation to 3 other types of QD that have not previously been studied in this way. We show that despite having a large difference in mean FSS at $F=0$, all samples display the same behavior. For all samples, $\gamma$ is constant with electric field and of a similar order of magnitude. Also, within a given sample $\gamma$ is the same for all QDs.

The observed linear variation in FSS is a general feature of self assembled QDs, but the rate of change of the FSS can be controlled by engineering the material properties of the QDs. We demonstrate that, in a given sample, the yield of QDs which are suitable for applications which require small $|s|$ depends not only on the distribution of $|s|$ amongst the QDs, but also on the rate at which the FSS can be tuned. In three of the four samples presented here we are able to minimize $|s|$ and identify QDs with sufficiently small $|s|$ to be suitable as sources of entangled photon pairs. In addition, avoided crossings are observed in the FSS of all QDs which can be tuned to their minimum $|s|$, including those emitting at around 1400 meV and those grown by the partially-capped island technique which were previously not thought to display such anticrossing behavior[10,11,13].

We begin by discussion of the growth procedure for each sample. All samples were grown inside a planar cavity with 14 periods below a λ/2 cavity and 4 periods above. The cavity consisted of a AlAs/GaAs superlattice with net 75% AlGaAs composition, within which a 10 nm quantum well (QW) was centrally positioned with the QD layer inside. The intrinsic region had a width of ~140 nm and was also centered on this spacer layer. The only difference between the samples is the growth conditions for the central QW and QD section. For sample A the QDs were grown on GaAs by depositing a single InAs layer for 72.5s at 470°C and then over growing with GaAs. This results in a low density of QDs at the cavity mode energy of 1324 meV. In sample B the InAs deposition time was reduced to 69s to result in a low density of QDs emitting in the 1383-1406 meV range, with the cavity thicknesses adjusted accordingly. It is well known that these QDs naturally have a low mean value of FSS at zero external field, and previous studies have shown that they are particularly promising as sources of entangled photon pairs[13]. This might be expected to be a result of improved symmetry of the confining potential, but we show here that despite a reduced mean FSS, the rate of tuning is broadly similar. Sample C consists of QDs grown in a single deposition in the center of a 33%AlGaAs layer, which increases the interface strain between the dot layer and cladding material. In this case the deposition time had to be increased to 74s at 470°C to account for the increased critical thickness required for QD self-assembly. A final sample, D, was prepared using the "partially capped island" growth technique to create short InAs QDs in GaAs[15, 12]. To achieve this, 100 seconds of InAs was deposited onto GaAs at a substrate temperature of 515°C, which was then capped with 2nm of GaAs. A 2 minute interrupt allowed the redistribution and desorption of indium before the growth of the rest of the structure is completed. Previous studies have shown this truncates the QD height[11].

Diode devices were then processed with a thin metal film on the sample surface patterned with apertures to allow small areas of the samples to be optically addressed. In all samples, doping extended into the 75%AlGaAs superlattice barrier leading to a built in voltage $V_{bi}$ = 2.2V. Thus we calculate the electric field which results from the application of a given voltage, $V$, as $F = (V-V_{bi})/d$, where the intrinsic region thickness is $d$ = 140 nm, for all diodes.

Photoluminescence was excited from the devices using a laser in the range 1464-1500 meV, and spectra recorded using a CCD/spectrometer system with a resolution of ~ 25 µeV. Polarised spectra, fit with Lorentzian lines as a function of linear polarization angle, allow us to determine sub-resolution shifts in the excitonic energy due to the FSS. In samples C and D the density of QDs was such that it was not possible to isolate a single QD, so a low excitation power was employed (to prevent excitation of the XX state). From each sample we have studied ~20 QDs selected for their brightness. Data showing the typical measurements of FSS and emission energy as a function of vertical electric field, $F$, are shown in Figure 1. A summary of the behavior of the sets of QDs studied from each sample is shown in Table I.

Several observations can be made from the data. Comparing samples A and B is it is clear that selecting QDs with emission energy closer to ~1400 meV, such as those in B, leads to a significant fall in the mean value and standard deviation of the FSS, as suggested by previous studies[13]. This translates to a larger fraction of QDs that can be tuned to a minimum FSS in B, relative to A. Such fractions can be quantitatively assessed via consideration of the rate at which the FSS can be tuned, $\gamma$, and the maximum magnitude of the field which can be applied without carriers tunneling out of the QDs[8], $F_{max}$. The tuning rates of samples A and B are $\gamma = -0.285 \pm 0.019$ μeV cm kV$^{-1}$ and $\gamma = -0.196 \pm 0.024$ μeV cm kV$^{-1}$, respectively. For sample A $|F_{max}|= 500$ kV cm$^{-1}$, but for sample B this is reduced to $|F_{max}|= 430$ kV cm$^{-1}$ due to the reduced confinement energies of QDs with higher emission energy. Thus, the range over which the FSS can be tuned in each sample, calculated from the product of $\gamma$ and $F_{max}$, is 145 μeV for A and 84 μeV for B. From the mean and standard deviation of the FSS at zero field, shown in table 1, the proportion of QDs in each sample which can be tuned to their minimum value is ~60% for A and ~90% for B.

The next device we study is C, which displays an increased mean FSS at $F=0$ of 644±134 μeV. This increase is expected, as the increased interface strain between the QDs and the cladding material increases the in-plane anisotropy of the QDs, on which $|s|$ depends[14], via asymmetric modification of the local strain field. Despite this increase of the in-plane anisotropy, the rate of FSS tuning, $\gamma$, is approximately the same as in sample A and so it was not possible to minimise $|s|$ in sample C. However, the remarkable similarity of the tuning rate in samples A and C suggests that $\gamma$ is robust to changes in the local strain field. This may be useful when designing QDs suitable for applications which require small |s|, as modifications to the strain field can be used to engineer other properties of the QDs without altering the available FSS tuning range.

The last device we study is from sample D. These "partially capped island" QDs have been widely studied within Schottky-diode heterostructures but their FSS has not been shown to change with electric field. Here we show that this is because these QDs have a reduced rate of tuning, which in conjunction with the limited value of $F_{max}$ for QDs within typical Schottky-diodes, resulted in a maximum change in FSS of only a few µeV in previous studies. In sample D we observe a mean FSS at *F = 0* of 27±16 µeV, which is smaller than that of samples A-C as expected, along with a value of γ that is the smallest of the samples studied here. The reduced rate of tuning is likely to be a direct result of the reduced height and indium content in these QDs, the former reason means it will be harder to shift the electron wavefunction to regions of different indium content and thus change the FSS[19] and the later has been predicted to result in a reduced ability to tune FSS[16]. Our experimental observations may provide some explanation for the unusual results presented in Seidl *et al*[15], where no preferred eigenstate orientation direction relative to the crystal lattice was observed. As can be seen from figure 2(d)-(f), far from its minimum |s| the preferred orientation of a QD's eigenstates aligns well with the ([110],[1-10]) crystalline axes; however as a QD is tuned through its minimum |s|, this orientation rotates through 90° due to coherent coupling between the two exciton eigenstates[5]. As the standard deviation of the FSS at *F=0* is a large fraction of the mean value it is plausible that a large proportion of the QDs are close to their minimum |s| at *F=0*. Consequently, the orientation of the eigenstates of each QD is strongly dependent on the strength of the coherent coupling between the two exciton eigenstates, which varies significantly between QDs, leading to an apparently random orientation of the eigenstates if measured at only one value of *F*.

Finally, we show plots of FSS and orientation angle for QDs in samples A, B and D, as the electric field is varied to minimise |s| (the FSS could not be minimised in C). In all samples the same coherent coupling behaviour is observed; the variation of |s| with electric field is described by a

hyperbola with a minimum value, $s_0$, which varies between QDs and a gradient away from minimum, $\pm\gamma$, which is a property of the QD growth conditions. Several theoretical studies[16, 17, 18, 19] have stated that a finite value of $s_0$ is to be expected for QDs grown on the [001] plane of GaAs, which this observation supports. Thus, the crucial parameter for determining the yield of QDs within a given sample which are suitable for small $|s|$ applications is the distribution of $s_0$, rather than the distribution of the FSS at $F=0$. If the QDs are engineered to have a FSS tuning range comparable to the mean FSS at $F=0$, and have a suitable distribution of $s_0$, a high yield of low $|s|$ QDs is possible without requiring the emission energy to be around 1400 meV or the use of the "partially capped island" technique.

In summary, we have shown that for the four types of self-assembled QD studied here the change of FSS with electric field can be parameterized by the FSS at $F = 0$, a minimum FSS magnitude $s_0$, and a linear tuning rate $\gamma$. This study may be of use to those theorists modeling how FSS changes with $F$, or experimentalists trying to increase the yield of QDs which can be tuned to a minimum value of FSS. Fruitful continuations of this study would be to measure QDs grown on other substrate orientations, such as [111] where the distribution of $s_0$ has a low mean value[20]; or positioned QDs, where the lateral shape may be controlled[21].

ACKNOWLEDGEMENT This work was partly supported by the EU through the IST FP6 Integrated Project QESSENCE. EPSRC provided support for MAP.

**References**

[1] O. Benson, C. Santori, M. Pelton, and Y,Yamamoto, Phys. Rev Lett., **84**,2513-2516 (2000).

[2] R. M. Stevenson, A. J. Hudson, A. J. Bennett, R. J. Young, C. A. Nicoll, D. A. Ritchie and A. J. Shields, Phys. Rev. Lett., 101, 170501 (2008).

[3] C. Santori, D. Fattal, M. Pelton, G. S. Solomon, and Y. Yamamoto, Phys. Rev. B **66**, 045308 (2002).

[4] R. M. Stevenson, R. M. Thompson, A. J. Shields, I. Farrer, B. E. Kardynal, D. A. Ritchie, and M. Pepper, Phys. Rev. B **66**, 081302 (2002).

[5] A. J. Bennett, M. A. Pooley, R. M. Stevenson, M. B. Ward, R. B. Patel, A. Boyer de la Giroday, N. Sköld, I. Farrer, C. A. Nicoll, D. A. Ritchie *et al*, Nature Physics 6, 947–950 (2010)

[6] A. Boyer de la Giroday, A. J. Bennett, M. A. Pooley, R. M. Stevenson, N. Sköld, R. B. Patel, I. Farrer, D. A. Ritchie and A. J. Shields, Phys. Rev. B, 82, 241301R.

[7] R. Trotta, E. Zallo, C. Ortix, P. Atkinson, J. D. Plumhof, J. Van den Brink, A. Rastelli, and O. G. Schmidt, Phys. Rev. Lett. 109, 147401 (2012)

[8] A. J. Bennett, R. B. Patel, J. Skiba-Szymanska, C. A. Nicoll, I. Farrer, D. A. Ritchie, and A. J. Shields, Appl. Phys. Lett. **97**, 031104 (2010).

[9] M. Ghali K. Ohtani, Y. Ohno and H. Ohno, Nat. Commun. 3:661 doi: 10.1038/ncomms1657 (2012).

[10] Z.R. Wasilewski, S. Fafard and J.P. McCaffrey, Journal of Crystal Growth, 201–202, 1131–1135 (1999).

[11] R. J. Warburton, C. Schäflein, D. Haft, F. Bickel, A. Lorke, K. Karrai, J. M. Garcia, W. Schoenfeld and P. M. Petroff, Nature 2000, 405, 926-929.

[12] J. M. Garcıa G. Medeiros-Ribeiro, K. Schmidt, T. Ngo, J. L. Feng, A. Lorke, J. Kotthaus and P. M. Petroff, Appl. Phys. Lett. **71**, 2014 (1997)

[13] R. J. Young, R. M. Stevenson, A. J. Shields, P. Atkinson, K. Cooper, D. A. Ritchie, K. M. Groom, A. I. Tartakovskii, and M. S. Skolnick, Phys. Rev. B **72**, 113305 (2005)

[14] M. Bayer, G. Ortner, O. Stern, A. Kuther, A. A. Gorbunov, A. Forchel, P. Hawrylak, S. Fafard, K. Hinzer, T. L. Reinecke, *et al.*, Phys. Rev. B. **65** 195315 (2002)

[15] S. Seidl, B.D. Gerardot, P.A. Dalgarno, K. Kowalik, A.W. Holleitner, P.M. Petroff,

K. Karrai and R.J. Warburton, Physica E **40**, 2153-2155 (2008).

[16] R. Singh and G. Bester, Phys. Rev. Lett. **104**, 196803 (2010).

[17] J. Luo, R. Singh, A. Zunger, and G. Bester, Phys. Rev. B **86**, 161302 (2012)

[18] R. Trotta, E. Zallo, C. Ortix, P. Atkinson, J. D. Plumhof, J. van den Brink, A. Rastelli, and O. G. Schmidt, Phys. Rev. Lett. **109**, 147401 (2012).

[19] G. W. Bryant, N. Malkova, J. Sims, arXiv:1212.4523 (2012).

[20] G. Sallen, B. Urbaszek, M. M. Glazov, E. L. Ivchenko, T. Kuroda, T. Mano, S. Kunz, M. Abbarchi, K. Sakoda, D. Lagarde *et al*, Phys. Rev. Lett. 107, 166604 (2011).

[21] P. Atkinson, S. Kiravittaya, M. Benyoucef, A. Rastelli, and O. G. Schmidt, Appl. Phys. Lett. 93, 101908 (2008).

**Table I.** Summary of the behavior of the exciton emission energy and fine structure splitting of QDs in each of the four sample types, A, B, C, and D. For all parameters, the value given is the mean average over all the QDs studied in the corresponding type. The columns contain the exciton emission energy, $E_0^X$; Stark shift parameters, z-dipole moment, $p^X$, and polarizability, $B^X$; fine structure splitting at zero electric field, $s(F=0)$; and the rate at which the fine structure splitting is tuned with electric field, $\gamma$.

| Type | $E_0^X$ (meV) | $p^X$ (meV cm kV$^{-1}$) | $B^X$ (meV cm$^2$ kV$^{-2}$) | $s(F=0)$ (μeV) | $\gamma$ (μeVcm kV$^{-1}$) |
|---|---|---|---|---|---|
| A. InGaAs/GaAs | 1327.3±7.4 | -(5.0±5.1) x 10$^{-3}$ | -(9.7 ± 0.9) x 10$^{-4}$ | 123 ± 67 | -0.285 ± 0.019 |
| B. InGaAs/GaAs | 1397.4±4.7 | -(6.5±2.8) x 10$^{-3}$ | (1.5±0.1) x 10$^{-4}$ | 48 ± 25 | -0.196 ± 0.024 |
| C. InGaAs/Al$_{0.3}$Ga$_{0.7}$As | 1426.8±3.7 | -(1.2±0.2) x 10$^{-3}$ | -(0.31±0.02)x10$^{-4}$ | 644 ± 134 | -0.306 ± 0.035 |
| D. InGaAs/GaAs (Partially capped islands) | 1344.2±5.4 | -(2.2±1.9) x 10$^{-3}$ | -(1.2±0.4) x 10$^{-4}$ | 27 ± 16 | -0.096 ± 0.036 |

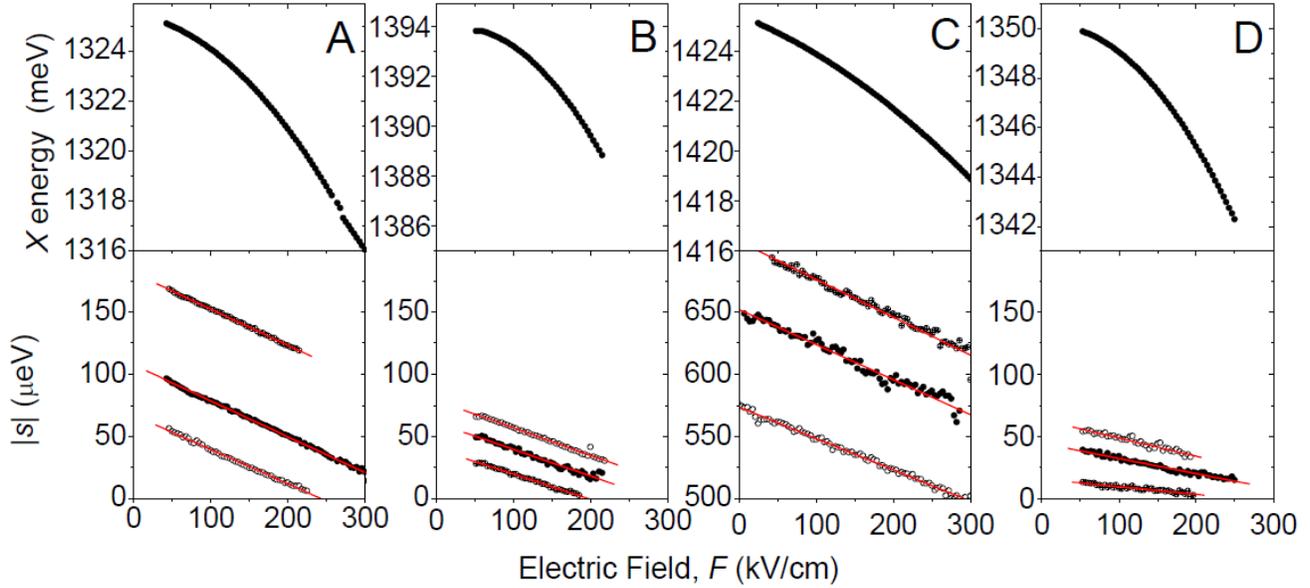

**Figure 1.** Electric field dependence of the exciton photon energy for (a) sample A, (b) sample B (c) sample C and (d) sample D showing the parabolic Stark shift of the exciton emission energy, from which the Stark parameters $p^X$ and $B^X$ are extracted, of a typical QD selected randomly from each sample. The lower panels show the magnitude of the fine structure splitting, $|s|$, for (e) sample A, (f) sample B, (g) sample C and (h) sample D illustrating the linear shift in $|s|$, from which $\gamma$ is extracted, of three typical QDs in each sample.

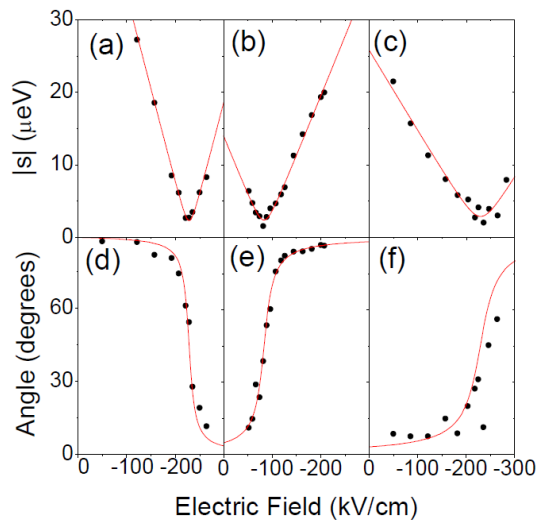

**Figure 2.** Observation of avoided crossings in the neutral exciton states of quantum dots tuned with electric field. (a), (b) and (c) magnitude of the fine structure splitting, |*s*|, in a single QD from samples A, B and D, respectively, as a function of electric field. In each case the minimum value of |s| is approximately 2 μeV and so is sufficient to show entangled photon emission. (d), (e) and (f) show the corresponding orientation of the eigenstates, relative to the [110] crystalline axis, as the QDs are tuned through their minimum |s|.